\documentclass[apj]{emulateapj}
\newcommand{\eg}{{\it e.g.,}}

\begin{document}
%===============================================================================
\title {Challenges for Precision Cosmology with X-ray and Sunyaev-Zeldovich Effect Gas Mass Measurements of Galaxy Clusters}
%===============================================================================
\author{Eric~J.~Hallman}
\affil{Center for Astrophysics and Space Astronomy, University of Colorado, Boulder, CO 80309}
\author{Patrick~M.~Motl}
\affil{Department of Physics and Astronomy,
Louisiana State University, Baton Rouge, LA 70803} 
\author{Jack~O.~Burns} 
\affil{Center for Astrophysics and Space Astronomy, University of Colorado, Boulder, CO 80309}
\author{Michael~L.~Norman}
\affil{Center for Astrophysics and Space Sciences, University of California-San Diego, 9500 Gilman Drive, La Jolla, CA 92093}
\email{hallman@casa.colorado.edu}

\begin{abstract}
We critically analyze the measurement of galaxy cluster gas masses, which
is central to cosmological studies that rely on the galaxy cluster
gas mass fraction. Using synthetic observations of numerically simulated
clusters viewed through their X-ray emission and thermal
Sunyaev-Zeldovich effect (SZE), we reduce the observations to obtain
measurements of the cluster gas mass.  We
utilize both parametric models such as the isothermal cluster model
and non-parametric models that involve the
geometric deprojection of the cluster emission assuming spherical
symmetry.  We are thus able to quantify the possible sources of
uncertainty and systematic bias associated with the common simplifying
assumptions used in reducing real cluster observations including
isothermality and hydrostatic equilibrium. We find that intrinsic
variations in clusters limit the precision of observational gas mass
estimation to $\sim$10\% to $1\sigma$ confidence excluding instrumental
effects. Gas mass estimates performed via all methods surprisingly
show little or no trending in the scatter as a function of cluster
redshift. For the full cluster sample, methods that use SZE profiles out to roughly the virial radius are the simplest, most accurate, and unbiased way to estimate cluster
mass. X-ray methods are systematically more precise mass estimators
than are SZE methods if merger and cool core systems are removed, but X-ray methods slightly overestimate (5-10\%) the
cluster gas mass on average. SZE methods are more precise and accurate than
X-ray methods at mass estimation if the sample is contaminated by
merging, disturbed, or cool core clusters. In fact we find that cool core
clusters in our samples are particularly poor candidates for observational mass
estimation, even when excluding emission from the core region. The
effects of cooling in the cluster gas alter the radial profile of the
X-ray and SZE surface brightness outside the cool core region, leading
to poor gas mass estimates in cool core clusters. Finally, we find that methods using a universal temperature profile
estimate cluster masses to higher precision than those assuming isothermality.  
\end{abstract}
%===============================================================================
%===============================================================================
\section{Introduction}
Measuring the apparent
change of cluster gas fraction with redshift produces constraints on the dark energy equation of
state, $w$ (e.g. \citet{pen}, \citet{sasaki}, \citet{rines}, \citet{vikh}).
Determining cluster
gas fractions requires high precision estimation of cluster gas mass
and total mass from
observables. In order for such methods to provide strong constraints
on cosmological parameters, cluster mass estimators must be accurate
to $\sim$10\% \citep{haiman}. Previous studies (e.g. \citet{evrard})
suggest that this level of precision may be difficult to achieve. The often used scaling relation
between cluster total mass and X-ray spectral temperature has relatively small
scatter compared to other methods, but the normalization is still uncertain to
$\sim$30\% (e.g. \citet{sand}). Additionally, using X-ray observations
and the assumption of hydrostatic equilibrium appears to lead to a bias
in cluster total mass estimates \citep{rasia06}. While it is necessary to estimate both
the cluster gas and total mass to determine the gas
fraction, this study aims first to determine the limiting accuracy of
various X-ray and Sunyaev-Zeldovich effect (SZE) observational methods
of cluster gas mass estimation. This analysis will help determine
whether cluster methods can be accurate enough to do precision
cosmology.  This study addresses the key
question: {\it What is the best way to measure cluster gas masses with
  precision in order to do cosmology?} 
 
High resolution X-ray or SZE observations of clusters coupled with assumptions
about the gas distribution lead to estimates of the gas mass in the
cluster dark matter potential well. The electron number density is
often assumed to fit a $\beta$ model,
\begin{equation}
  n_{e}(r) = n_{e0}\left(1 + \left(\frac{r}{r_{c}}\right)^{2}\right)^{-3\beta/2}.
\end{equation}
Fitting an observed X-ray or SZE profile to these projected $\beta
$ model X-ray surface brightness and SZE $y$ parameter distributions
results in a description of the density distribution, which can be
integrated to obtain the gas mass. The difference in dependence on gas
density and temperature of X-ray emissivity and the SZE $y$ parameter
makes the combination of these two methods of observation very
powerful. Because of this difference, the observability of clusters
via each method is affected differently by the impact of central physics including
radiative cooling and feedback mechanisms, as well as the transient
boosting of surface brightness and spectral temperature generated
during merging events. Thus these two methods not only select a
different sample of clusters, but combined SZE/X-ray observations of
individual clusters allow one to extract the density and temperature
of the gas without relying on X-ray spectral temperatures.

Recent numerical (e.g. \citet{loken}) and observational
  studies (e.g. \citet{vikh}) suggest that the cluster gas is not
  isothermal, but fits more closely to a universal temperature profile
  (UTP), for this study written as
\begin{equation}
  T(r) = \langle T \rangle_{500}T_{0}\left(1 + \left(\frac{r}{\alpha r_{500}}\right)^{2}\right)^{-\delta},
\end{equation}
where $ \langle T \rangle_{500}$ is the average
temperature inside a projected overdensity radius of $r_{500}$. Here
the subscript indicates that the ratio of the average overdensity inside this
radius with respect to the comoving universal mean is equal to 500. $T_0$,
$\alpha$, and $\delta$ are parameters whose mean value is measured
from the entire cluster population at some redshift. We perform a
deprojection of the X-ray and SZE profiles and use this additional
assumption about the gas temperature to calculate the density profile
and the mass. 

With observations of both X-ray and SZE for a particular
cluster, an observer can deproject the surface brightness of each
simultaneously to determine the gas density.  This method is
particularly powerful due to the difference in dependence of X-ray and
SZE emission on density and temperature.  With combined observations,
one can in principle determine the cluster gas mass with no
assumptions about the cluster temperature profile, using only the
weaker assumption of spherical symmetry in the
deprojection. In this study, we also perform such a combined
deprojection of X-ray and SZE profiles to obtain the gas density
radial profile with no assumption about the temperature profile.

We examine the effect of the assumption of the gas temperature
  and density profile on the determination of the true cluster gas
  mass. Using adaptive mesh hydro/N-body
  cosmological simulations, we have extracted clusters with $M \geq
  10^{14} M_{\odot}$ out to a redshift of z = 2.  This sample has
  $\sim$100 such clusters at z=0 and $\sim$10 clusters at z = 2. For
  each cluster, we have fit the X-ray and SZE surface brightness profiles to those produced by a best-fit
  $\beta$-model gas distribution. We also perform a direct geometric deprojection of the
  SZE and X-ray surface brightness. We analyze them separately assuming the gas
  temperature follows a UTP, as well as jointly with no assumptions
  about the temperature distribution to determine a density profile and
  thus the mass. The joint method requires spatially resolved X-ray and SZE
  profiles for each cluster and assumes only spherical symmetry. We also compare profile
  methods with cluster X-ray and SZE scaling relations for accuracy in
  mass determination. 

The outline of this work is as follows: Section 2 describes the
numerical simulations performed, Section 3 details the methods of
observational cluster gas mass estimation used, Section 4 contains
results of that mass estimation, and Section 5 states conclusions.
\section{Simulations}
The simulations described here use the cosmological hydro/N-body
adaptive mesh refinement (AMR) code {\it Enzo} (\citet{enzo};
  http://cosmos.ucsd.edu/enzo) to evolve both the dark matter and
  baryonic fluid in the clusters utilizing the piecewise parabolic
  method (PPM) for the hydrodynamics. We achieve a peak resolution of
  15.6$h^{-1}$kpc with seven levels of refinement. Refinement of high
  density regions is done as described in \citet{motl04}. We assume a
  concordance $\Lambda$CDM cosmological model with the following
  parameters: $\Omega_b = 0.026, \Omega_m = 0.3, \Omega_\Lambda = 0.7,
  h = 0.7,$ and $\sigma_8 = 0.928$. 
  
  The simulations used for this study were briefly described in
  \citet{motl05}. The
  same four sets of simulation data are used here,
  including progressively more baryonic physics.  The baseline
  simulation was run with purely adiabatic physics for the baryons.
  The other three simulations add radiative cooling, star formation,
  and star formation with a
  moderate amount of thermal feedback from supernovae using the
  \citet{cen} algorithm.  The radiative cooling is calculated via a
  Raymond-Smith model for the X-ray emission.
\section{Methods of Gas Mass Estimation}
We use the catalog of AMR refined clusters identified in each simulation
volume which was described in \citet{motl05}. Each catalog from
the four simulations contains $\approx$ 100 clusters from the current
epoch in the mass range $10^{14} M_{\odot} \leq M \leq 2 \times
10^{15} M_{\odot}$ ($\sim 15$\% of Virgo mass to $\sim$Coma mass) and also
contains all clusters above $10^{14} M_{\odot}$ for a series of twenty
redshift bins out to $z = 2$. This corresponds to roughly 1500
clusters in the interval $0 < z < 2$ in each catalog. 

Our objective was to determine the reliability of observational cluster mass
estimation methods. We examine the relative accuracy of these methods
with the cluster observables in their simplest form, uncontaminated by
instrumental effects.  This allows us to determine the limiting
accuracy of any particular method for a realistic,
mass-limited set of clusters in a cosmological environment. 
In order to perform a mass estimation for these clusters in a method
similar to that done observationally, we first generate projected maps
of each cluster volume of their X-ray surface brightness and
SZE y parameter.  The X-ray surface brightness at any point in the
projected map is generated by
integrating along the line of sight the bremsstrahlung
emissivity
\begin{equation}
S_X = \frac{1}{4 \pi (1+z)^4} \int n_e(l) n_H(l) \Lambda(T(l)) dl
\end{equation}
where $\Lambda$ is the emission function for the cluster gas.
Generally this will include thermal bremsstrahlung emission in
addition to spectral line emission. For most of this study, $\Lambda$ is simply
\begin{equation}
\Lambda(T) = \Lambda_0 T^{1/2},
\end{equation}
which is a simple approximation to the X-ray emission representing the thermal bremsstrahlung
which dominates massive clusters with high virial temperatures ($>$ 2
keV).  However, for a subset of the clusters evolved to z=0, we have
calculated the thermal emissivity via a Raymond-Smith model assuming a
metallicity of 0.3 of the solar value \citep{raymond}. 
 
The SZE Compton $y$ parameter is calculated on each line of sight as \citep{sz}, 
\begin{equation}
y = \int \sigma_T n_e(l) \frac{k_b T(l)}{m_e c^2} dl.
\end{equation}
One additional map that is very helpful in mass estimation is that of
the spectral temperature.  In this study we have used the emission
weighted temperature as
\begin{equation}
T_{ew} = \frac{\int [n_e n_H \Lambda(T)] T dl}{\int n_e
  n_H \Lambda(T) dl}.
\end{equation}
For our subset of z=0 clusters for which we have generated
Raymond-Smith emissivities to calculate the X-ray surface brightness,
we have used a projected ``spectroscopic-like temperature'' ($T_{sl}$ as in
\citet{rasia}). This value for the projected temperature more closely
approximates the spectral temperature of the cluster that would be
determined from the X-ray spectrum.  The value of $T_{sl}$ used here is 
\begin{equation}
T_{sl} = \frac{\int n_e n_H T^a/T^{1/2} dl}{\int n_e n_H T^a/T^{3/2}
  dl},
\end{equation}
where a=0.75 is a fitted parameter from \citet{rasia}.

From the simulation data, we are able to generate these three maps at
three orthogonal orientations of the cluster. We then have spatially
resolved images and spectral temperatures for each cluster.  From
these maps, we can use typical observational techniques to estimate
the cluster gas mass, and compare to the true mass of each cluster on
the simulation grid.  For this study we have restricted ourselves to
methods which require only radial profiles of the surface brightness
and average spectral (emission-weighted) temperature. X-ray methods
that use spectral temperature profiles require much deeper
observations in order to achieve good photon statistics at large radius,
and therefore are costly observations for large cluster samples.  
\subsection{Beta Models}
As in Eq. 1, the gas density
distribution is often described as a beta model \citep{caval}.  A beta model density
distribution results in an X-ray projected radial surface brightness
distribution of the form
\begin{equation}
S_X(r) = S_{X0} \left(1 + \left(\frac{r}{r_c}\right)^2 \right) ^
{\frac{1}{2} - 3\beta} 
\end{equation}
where
\begin{equation}
S_{X0} \propto n_{e0}^2 \langle T \rangle^{\frac{1}{2}}.
\end{equation}
This proportionality holds in the bremsstrahlung limit, but the
temperature dependence is weaker in a fixed X-ray band \citep{mohr99}.
In the above relation, $n_{e0}$ is the central density normalization
of the beta model, and $\langle T \rangle$ indicates the average
spectral temperature inside a given radius.
Similarly for the SZE, a beta model density distribution results in a
projected radial distribution of the Compton $y$ parameter 
\begin{equation}
y(r) = y_0 \left(1 + \left(\frac{r}{r_c}\right)^2\right)^{\frac{1}{2}
  - \frac{3\beta}{2}}
\end{equation}
where 
\begin{equation}
y_0 \propto n_{e0} \langle T \rangle.
\end{equation}
Note particularly the stark difference in dependence on density and
temperature for the X-ray surface brightness and Compton $y$
parameter. 
This difference contributes to the subsequent variation in the quality
of mass estimates using X-ray or SZE methods.

Bolstered by earlier observational evidence, the hot gas in clusters is commonly assumed to be 
isothermal (\eg \space \citet{shim}). In the isothermal case, the
$\beta$-model has a physical interpretation, namely it is the
density profile which approximates an isothermal King model
\citep{king66,king72}. By fitting isothermal $\beta$-model relations to radial profiles of X-ray surface
brightness and Compton $y$ parameter, respectively, we obtain a value for
$\beta$, as well as a normalization of the profile. We use an
average spectral temperature (in this case emission weighted
temperature) to then calculate a value for $n_{e0}$. Integrating the
density profile then results in an estimate of the gas mass for each
cluster. 
\subsection{Universal Temperature Profile Methods}
It has been shown both in our simulations \citep{loken} and
in X-ray observations \citep{vikh} that hot gas in many clusters is not
isothermal, but follows a universal temperature profile (UTP) where
temperature declines with radius.
We have also performed mass estimates using the assumption of a UTP in
the cluster gas.  The method of mass estimation in this case involves
a geometric deprojection of the cluster radial surface
brightness or Compton $y$ parameter distribution. This deprojection is
performed in a similar way to previous studies, via the method of
\citet{kriss}, originally described in \citet{fabian}, seen more recently in \citet{buote00}. Under the assumption of spherical symmetry, the emissivity as
a function of radius in three dimensions can be recovered by
recognizing that the luminosity of each projected annulus of the
cluster results from contributions from spherical shells which overlap
one another. 

For the X-ray, the emissivity profile can be converted to a density
profile, provided one knows the temperature in each shell.  This
information is provided by the UTP, the normalization of which is set
by the value for the average projected emission weighted temperature
inside $r_{500}$. Similarly for the SZE, one can calculate the value
of Compton $y$ per unit volume in each shell, and with a temperature
from UTP, extract the density in each shell.  Then, in either the case of
X-ray deprojection or SZE deprojection, we simply add up the mass in
the shells out to some fiducial radius to get the total gas mass. In
the X-ray method, one also has the option of simultaneously
deprojecting the spectral temperature to get a three-dimensional
temperature profile, calculating the mass in that way. This method is limited due to the necessity of a large number
of photon counts in bins at large radius. 
\subsection{Joint SZE/Xray Methods}
Ideally, if one has both an X-ray and SZE profile of each cluster, one need not make any assumption about
cluster temperature.  Due to the different dependence of SZE and X-ray
emissivity on density and temperature, a profile of each can be used
to eliminate the temperature dependence (e.g. \citet{patel}). With deprojected X-ray and SZE profiles, we are able to
directly calculate the density profile with no reliance on UTP or any
information about temperature whatsoever.  We have also calculated
mass estimates in this way for each cluster in our simulated dataset. 
\subsection{Scaling Relations}
For completeness, we should mention that cluster total mass is often
deduced from X-ray scaling relations, and can be deduced in a similar
way from SZE scaling relations. Scaling relations between various
X-ray observables and derived quantities are noted to follow well
behaved power laws. Similar scaling relations between physical
quantities in clusters are predicted from theory. A typically used
scaling relation relates average X-ray spectral temperature to cluster
total mass (e.g. \citet{finoguenov,vikh05}), and X-ray
spectral temperature to cluster gas mass (e.g. \citet{mohr99, vikh99}).  Observed scaling relations are noted
to differ from these simple expectations for low mass clusters, which
has been interpreted as an entropy floor \citep{caval98,bial,bryan}. 

As we have shown in previous work \citep{motl05}, there is
also a tight correlation between integrated SZE $y$ parameter and the
cluster total mass. In this study, we show these results in comparison
to the other types of estimates detailed above. The scaling relations
used in this study to estimate mass are best-fit relations for the
simulated cluster catalogs.
\begin{figure*}
%\epsscale{1.05}
\plotone{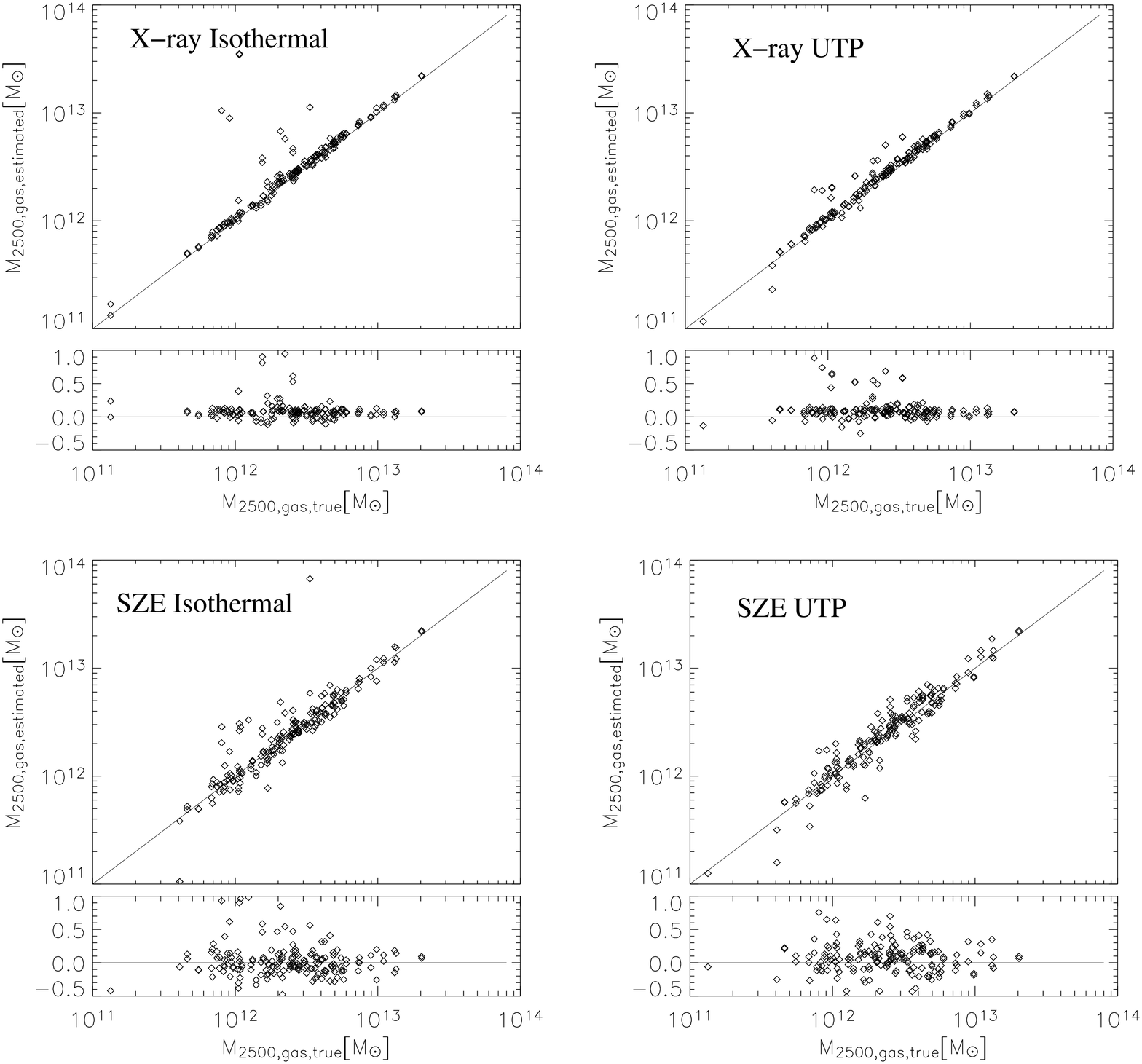}
\caption{Top: Left panel
  shows a plot of estimated gas mass to true gas mass inside
  $r_{2500}$ ($\sim 0.5h^{-1}$ Mpc) for each cluster
  at z=0 in the simulation. Mass is estimated using the X-ray surface
  brightness distribution and fitting to a
  $\beta$-model. Right panel is similar plot for deprojection of X-ray
  assuming a UTP model for temperature. Bottom: Left panel is similar
  mass plot for fitting the SZE $y$
  parameter distribution to a $\beta$-model; right panel is similar
  but for SZE deprojection assuming a UTP for temperature. Line in each
  case represents $M_{est} = M_{true}$. Bottom plot in each panel is
  log base e deviation of mass ratio $M_{est}/M_{true}$ from 1.0.} 
\label{mass_est_1}
\end{figure*}

\section{Results}
For each cluster in each of the four simulation catalogs in each
redshift interval, we have calculated estimated and true masses via
the above methods.  We calculate gas mass estimates from two randomly
selected orthogonal projections using each method, and for each of
three radii, $r_{2500} (\sim0.5h^{-1}Mpc), r_{500}(\sim1.0h^{-1}Mpc)$ and $r_{200}(\sim1.5h^{-1}Mpc)$.  In each case the
subscripted number indicates the relative average overdensity of the
gas inside that radius with respect to the comoving mean density at
each redshift.  We have calculated each overdensity radius directly
from the three-dimensional simulation data, because here we are
interested in understanding the best case scenario for cluster mass
estimation. In real clusters, the radius corresponding to each
overdensity must be derived from the observations, thus contributing
additional instrumental errors to the mass estimate. The radii are chosen in the three cases to match,
respectively, the approximate cluster overdensity radii which can be
imaged by Chandra ($r_{2500}$) for many clusters, approximate
overdensity radius which can be imaged by XMM-Newton ($r_{500}$), and
the approximate virial radius of the cluster in a $\Lambda$CDM universe
($r_{200}$).  In all cases we have excised cool cores from the
calculation of spectral temperature, projected quantities and mass
estimates. We cut the profiles at the 
point where the projected temperature profile peaks, so that the region excluded extends from the center to the point where the 
temperature profile begins to decrease again. Observationally, the
cool core 
could also be excluded by excising the strongly peaked
surface brightness region from the radial profile.
%===============================================================================
\begin{figure*}
\plotone{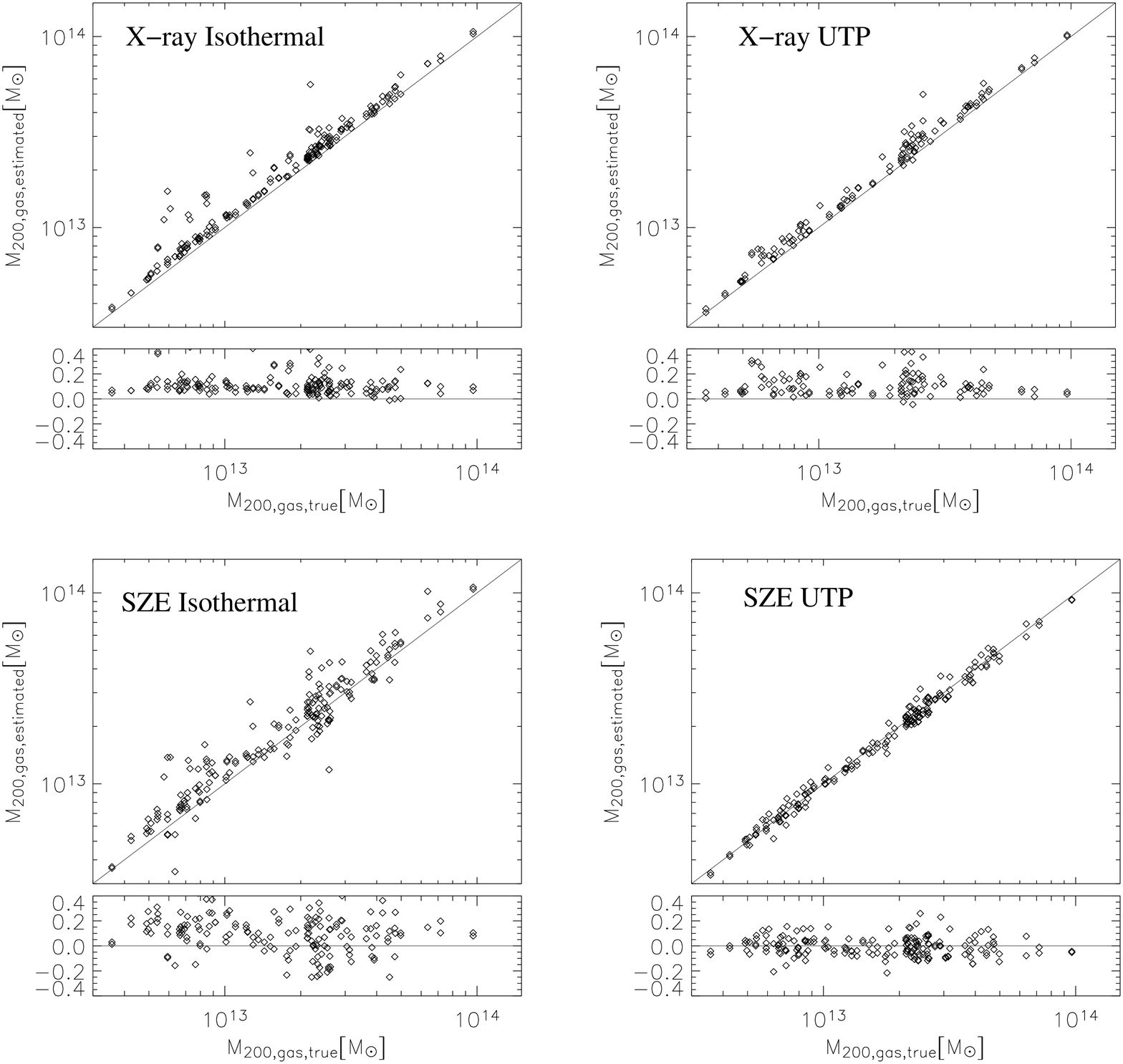}
\caption{Top: Left panel
  shows a plot of estimated gas mass compared to true gas mass inside $r_{200}$ for each cluster
  at z=0 in the simulation. Mass is estimated using X-ray surface
  brightness distribution and fitting to a
  $\beta$-model. Right panel is similar plot for deprojection of X-ray
  assuming a UTP model for temperature. Bottom: Left panel is a
  similar mass plot for fitting SZE $y$
  parameter distribution to a $\beta$-model; right panel is similar
  but for SZE deprojection assuming a UTP for temperature. Line in each
  case represents $M_{est} = M_{true}$.  Bottom plot in each panel is
  log base e deviation of mass ratio $M_{est}/M_{true}$ from 1.0.} 
\label{mass_est_2}
\end{figure*}

%=====================================================================
\begin{figure*}
\plotone{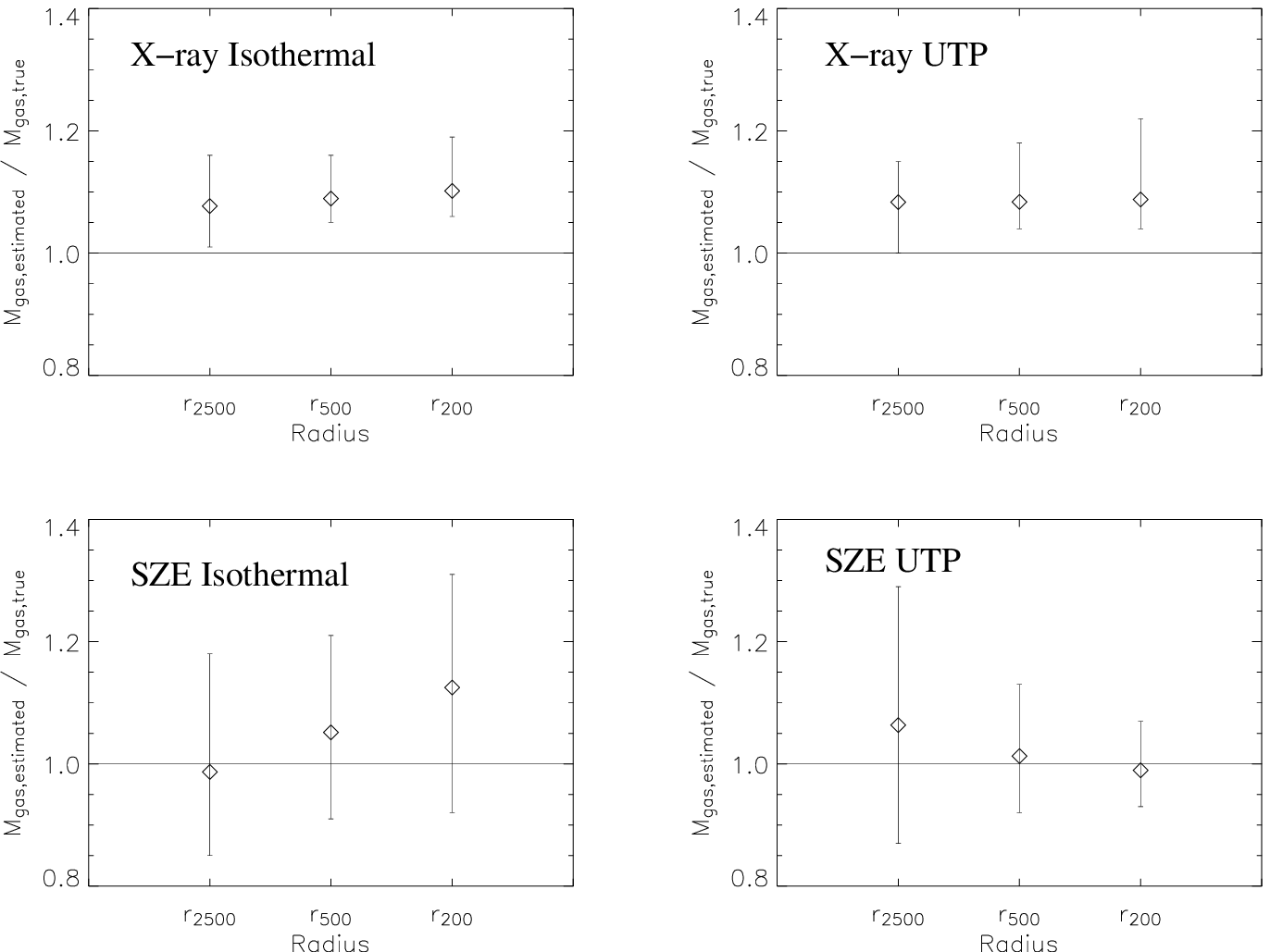}
\caption{Top: Left panel
  shows median ratio of estimated gas mass to true gas mass inside
  various radii as indicated for all clusters above $10^{14} M_{\odot}$ 
  at z=0 in the simulation for X-ray isothermal mass estimate. Error bars are 1$\sigma$ scatter.
  Right panel is similar plot for deprojection of X-ray assuming a UTP model for temperature. Bottom: Left panel is mass ratio for fitting SZE $y$
  parameter distribution to isothermal model, right panel is similar
  but for SZE deprojection assuming a UTP for temperature. Line in each
  case represents $M_{est} = M_{true}$. }
\label{radii} 
\end{figure*}

%===========================================================================
\subsection{Mass Estimates for z=0 Clusters}
First, we show results for the simulated cluster catalog which were
generated including all of the following: adiabatic physics, radiative cooling, star
formation, and energy feedback from stars (hereafter SFF).  Figure \ref{mass_est_1}
shows the true mass of the cluster plotted against the mass estimated
via 4 methods for clusters at z=0 in the simulation.  The top two
panels show X-ray mass estimates, and the bottom show estimates using
synthetic SZE data.  The left column plots result from use of the
$\beta$ model, and the right from deprojection and UTP. All
four plots estimate mass inside $r_{2500}$.  At first glance, it is
clear that the X-ray methods have a smaller scatter about the median
value than do the SZE methods.  Using UTP for the X-ray data appears
to eliminate some of the outliers generated using the assumption of
isothermality. Otherwise, there appears to be little difference in the
accuracy of the mass estimation whether using isothermal or UTP
methods in either X-ray or SZE case.  However, in Figure
\ref{mass_est_2}, we show mass estimates for the same clusters using
data which extends to a radius of $r_{200}$. Here,
using a UTP generates a substantial improvement in the scatter of the
mass estimates with SZE images.  No such major improvement occurs with
X-ray images.  Tables \ref{scatter2500} and \ref{scatter200} show the
relative scatter for these methods for $r_{2500}$ and $r_{200}$
respectively. This result is consistent with our previous work
\citep{motl05}, which indicates that when SZE data are available to
larger radii, cluster core effects are smoothed out, resulting
in better correlation of the signal with mass. Since X-ray emissivity
is proportional to $n^2$, it is always core dominated, and so
shows no improvement with inclusion of data from larger
radii. 

This issue is underscored by the plots in Figure \ref{radii}. These
show for the same methods as Figures \ref{mass_est_1} and
\ref{mass_est_2} the trend of the median value of the ratio of
estimated to true mass with limiting radius. Error bars are $1\sigma$
limits.  They show that the
upper error bars at all radii for the X-ray UTP method are smaller
than for the isothermal method.   The X-ray methods gain no
significant improvement if we have data out to larger radii.  This
trend indicates that X-ray mass estimates performed from observational
data which can give radial profiles to a fraction of the virial radius
are as good as those done with profiles which extend to the virial
radius. 

We also see that for the SZE
method, use of the UTP results in smaller errors when mass is
estimated to larger radii, whereas the isothermal method gives no
improvement regardless of radius used. SZE estimates, by
contrast to X-ray, have significantly reduced scatter when Compton $y$ parameter
profiles can be determined out to the virial radius of the cluster,
provided we use a realistic model for the temperature
profile. Isothermality is a better approximation at small radii to the
true temperature profile, but it breaks down at larger radii. 

Also significant in these plots is the apparent bias of the
median value of the X-ray mass estimates. It
is clear that both isothermal and UTP methods have a a median value
which is biased high, outside  the $1\sigma$ error range for all but the
smallest radius of estimation. This bias is approximately 5-10\%, and
results from the greater sensitivity of X-ray emissivity to transient
boosting by substructure, cooling, and mergers which are ubiquitous in
clusters. This result is consistent with previous work
\citep{mohr99,math99}, which claim a bias of cluster gas mass of
10-12\% from simulated clusters. We will discuss this
in more detail in Section 4.4.
\begin{table}
\caption{Scatter in $M_{est}/M_{true}$, $r_{2500}$, z=0}
\label{scatter2500}
\begin{tabular}{cccccc}
\tableline\tableline
Method & Model & Median & Mean & +1$\sigma$ & -1$\sigma$ \\
\tableline
X-ray & UTP & 1.08 & 1.13 & 1.15 & 1.00\\
SZE & UTP & 1.06 & 1.08 & 1.29 & 0.87\\
X-ray & Isothermal & 1.08 & 5.66 & 1.16 & 1.01\\
SZE & Isothermal & 0.99 & 1.18 & 1.18 & 0.85\\
Joint & Geometric & 1.09 & 1.11 & 1.16 & 1.04\\
\tableline
\end{tabular}
\end{table}

Lastly, in Table \ref{scatter2500} and Table \ref{scatter200} we list the scatter from the joint
X-ray/SZE geometric deprojection, which requires no assumptions about
the gas temperature.  It provides an estimate of the mass of similar
accuracy to X-ray data alone assuming a UTP for the clusters.

It is clear from Tables \ref{scatter2500} and \ref{scatter200} that if
measurements of cluster SZE profiles out to $r_{200}$ are available,
then the SZE method of cluster gas mass estimation using a UTP produces the
smallest scatter of the methods examined here.  In addition, there is
no bias in the median values within the errors, which is not true of
the X-ray methods.  For the full sample of clusters in the catalog,
UTP SZE method is the most accurate at estimating gas masses. 
\begin{table}
\caption{Scatter in $M_{est}/M_{true}$, $r_{200}$, z=0}
\label{scatter200}
\begin{tabular}{cccccc}
\tableline\tableline
Method & Model & Median & Mean &  +1$\sigma$ & -1$\sigma$ \\
\tableline
X-ray & UTP & 1.09 & 1.13 & 1.22 & 1.04\\
SZE & UTP & 0.99 & 0.99 & 1.07 & 0.93\\
X-ray & Isothermal & 1.10 & 1.17 & 1.19 & 1.06\\
SZE & Isothermal & 1.12 & 1.15 & 1.31 & 0.92\\
Joint & Geometric & 1.12 & 1.16 & 1.26 & 1.05\\
\tableline
\end{tabular}
\end{table}
%=======================================================================

For a randomly selected sub-sample of 26 clusters at z=0 ($\sim 25\%$ of
the sample), we have generated projections of the X-ray surface
brightness via the Raymond-Smith model assuming a metallicity of 30\%
of the solar value.  We also calculated the spectroscopic-like
temperature ($T_{sl}$) maps for each of the projections in this
sub-sample. The purpose of this exercise was to determine if there are
significant differences in the cluster mass estimation when using
realistic X-ray model emission instead of a simple bremsstrahlung
model. The analysis of each set of images is identical to that
described for the full sample of clusters.  

Table \ref{rs_emiss} summarizes the results of mass estimation using
an isothermal $\beta$-model method for the sub-sample of cluster
projections out to a radius of $r_{500}$ compared to the results for
the full sample.  We have done the sub-sample analysis for two X-ray
bands, 0.5-2.0 keV and 2.0-8.0 keV. The table shows that there is
a reduction from the simple bremsstrahlung analysis in
the precision of mass estimation in both the soft and hard band. 
The $1\sigma$ scatter has gone up by a factor of roughly 2-3. It is
interesting to note that the mean value for the bolometric
bremsstrahlung case is much higher than for the fixed X-ray band
case.  This is largely a result of the increased boosting of cool core
clusters in the bolometric method, which we will describe in more detail
later. It appears from our analysis that the simple bremsstrahlung
case represents a lower limit on the scatter.    
\begin{table}
\caption{X-ray Estimates of $M_{est}/M_{true}$, $r_{500}$, z=0}
\label{rs_emiss}
\begin{tabular}{ccccc}
\tableline\tableline
X-ray Energy Band & Median & Mean &  +1$\sigma$ & -1$\sigma$ \\
\tableline
Bremsstrahlung & 1.09 & 2.59 & 1.16 & 1.05\\
0.5-2.0 keV & 1.09 & 1.18 & 1.27 & 1.02\\
2.0-8.0 keV & 1.17 & 1.24 & 1.40 & 1.06\\
\tableline
\end{tabular}
\end{table}
%=======================================================================

\subsection{Mass Estimates as a Function of Redshift}
\begin{figure*}
\plotone{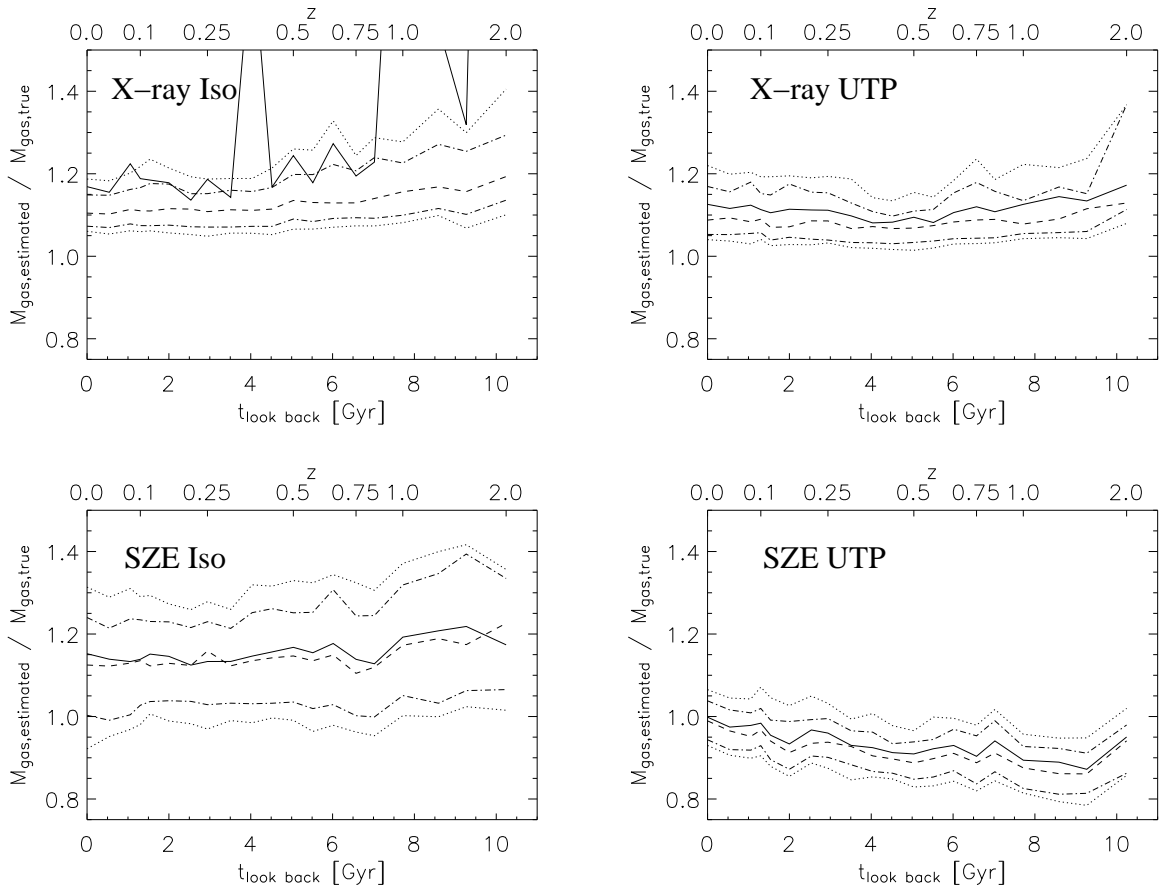}
\caption{Mass Estimation as a function of redshift in the simulation cluster catalog. Top: Left panel
  shows mean (solid line) and median (dashed line) ratio of estimated
  gas mass to true gas mass inside $r_{200}$ for isothermal X-ray
  method. Dot-dashed line represents range within which the middle
  50\% of clusters are estimated, dotted is for 1$\sigma$ limits. Right panel is similar plot for deprojection of X-ray assuming a UTP model for temperature. Bottom: Left panel is mass ratio for fitting SZE $y$
  parameter distribution to isothermal model, right panel is similar
  but for SZE deprojection assuming a UTP for temperature. }
\label{as_f_z}
\end{figure*}

We have also examined the evolution of our target mass estimation
techniques with redshift, back to z = 2.  For SZE methods, this is
particularly important due to the redshift independence of the flux of
the SZE,
since many clusters at redshifts above z=1.0 will be observable with
current and upcoming instruments. Figure \ref{as_f_z} shows plots of
the trend of the ratio of estimated mass to true mass for each of the
four methods and cluster sample described above, but as a function of
simulation redshift and for $r_{200}$. It is obvious that the relative
scatter of mass estimates is to first order redshift independent. This
is somewhat surprising, considering we have not eliminated any
obviously merging or disturbed clusters from the sample, and that
these should represent a higher fraction of the total sample at higher
redshift. This trend suggests that a higher major merger rate in
clusters as a whole does not
necessarily indicate a poorer overall quality of gas mass
estimates. The effect of continuous infall of subgroups and diffuse
matter over the lifetime of clusters may be as important a contributor
to scatter in gas mass estimates as the frequency of major mergers. 

It is obvious from the solid line indicating the mean value for the
X-ray isothermal $\beta$-model method that there are some outliers which drive the mean very high at two
redshifts in particular, which do not have an analogous effect in the
UTP model method. However, in the X-ray, using UTP does not
significantly improve the overall scatter. This is likely due to the
weak temperature dependence of X-ray emissivity. For the SZE, however, it is obvious that at all redshifts,
one gains an advantage by using the UTP.  The stronger dependence on
temperature of the SZE means that poor assumptions about temperature lead
to larger errors in mass estimation.  

\begin{figure}
\epsscale{1.25}
\plotone{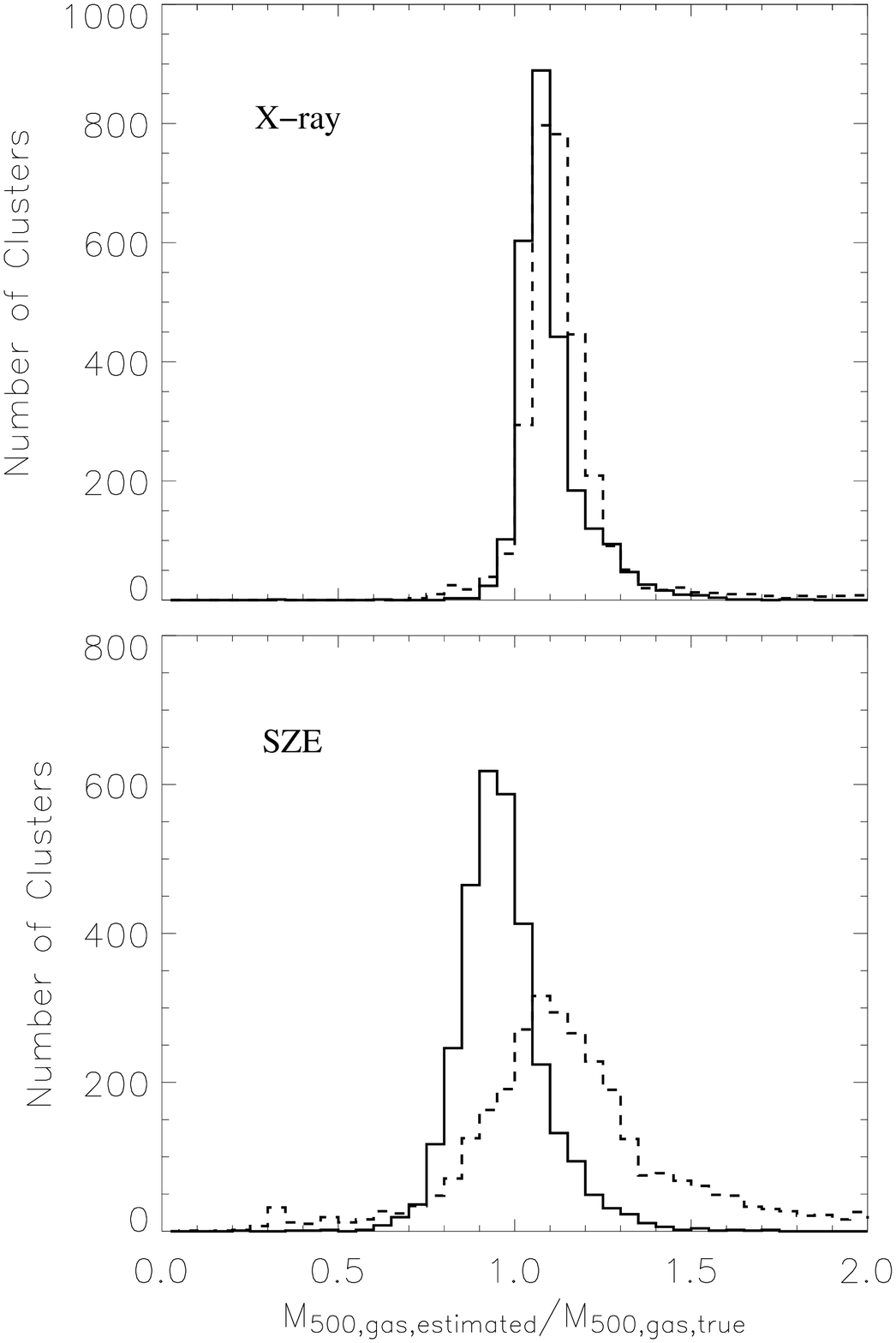}
\caption{Top: Distribution of X-ray estimated to true gas mass ratios
  inside $r_{500}$ for
  all cluster projections in the simulation of clusters over $10^{14}
  M_{\odot}$ from z=0 to z=2. Dashed line is for
  $\beta$-model method, solid is for UTP method. Bottom: Distribution
  of mass estimates for same cluster sample as top panel, but for SZE
  methods.}  
\label{ensem}
\epsscale{1.0}
\end{figure}

\begin{figure*}
\plotone{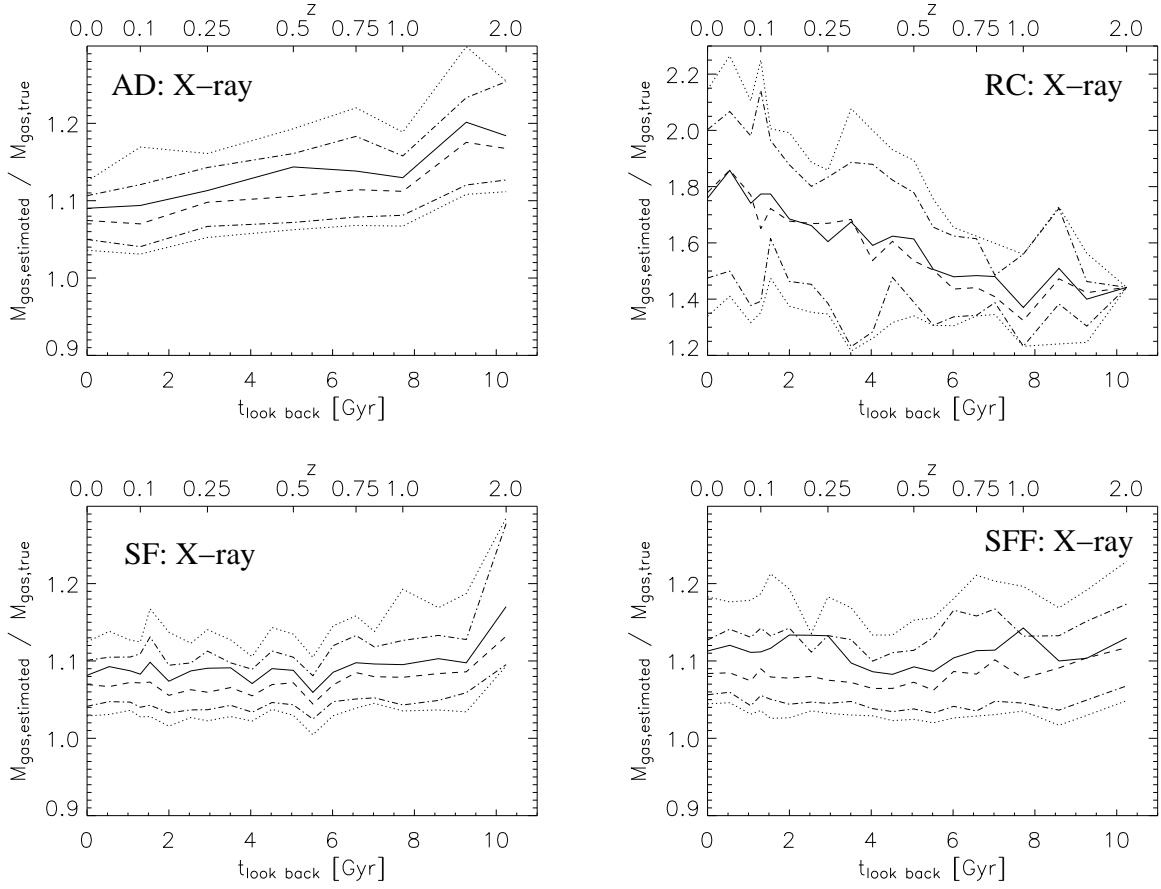}
\caption{Mass Estimation as a function of redshift in the simulation cluster catalogs. Top: Left panel
  shows mean (solid line) and median (dashed line) ratio of estimated
  gas mass to true gas mass inside $r_{500}$ for X-ray UTP method for
  the adiabatic catalog (AD). Dot-dashed line represents range within which the middle
  50\% of clusters are estimated, dotted is for 1$\sigma$ limits. Right
  panel is similar plot but for the catalog of clusters including
  adiabatic physics and radiative cooling (RC). Note that for RC plot,
  vertical scale differs from the other panels. Bottom: Left panel is mass
  ratio for the SF catalog, right panel is similar to the others but
  for the SFF catalog. }
\label{bary_phys}
\end{figure*}

%==========================================================================
\begin{figure*}
\plotone{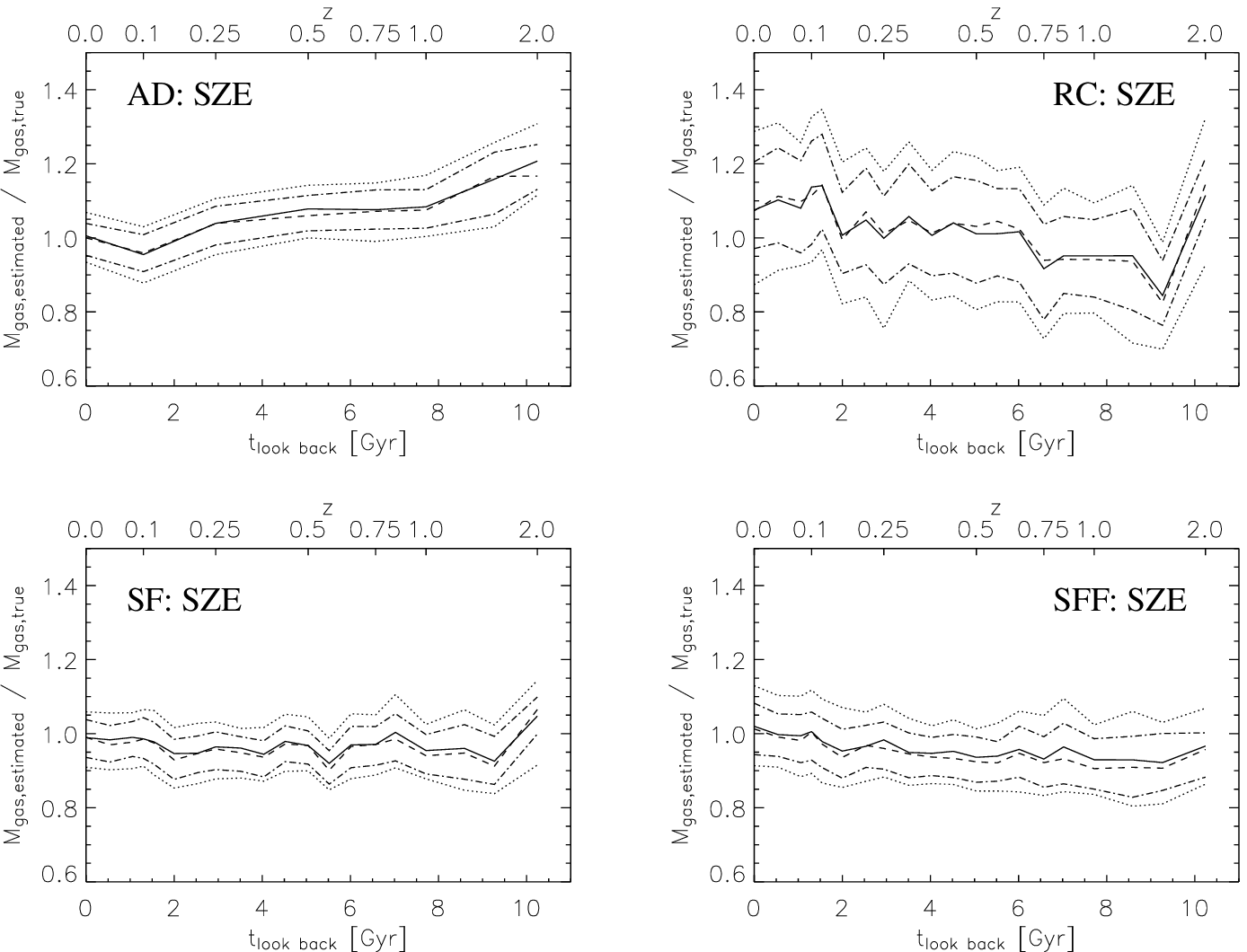}
\caption{Similar to Fig. \ref{bary_phys}, but for mass estimation
  using the SZE UTP method. Each plot shows mean (solid line) and median (dashed line) ratio of estimated
  gas mass to true gas mass inside $r_{500}$. Dot-dashed line represents range within which the middle
  50\% of clusters are estimated, dotted is for 1$\sigma$ limits. Labels
  above plots indicate baryonic physics included in each catalog. }
\label{bary_phys_2}
\end{figure*}
In Figure \ref{ensem} we have plotted the histograms indicating the
ratio of estimated to true cluster gas mass for the full mass limited SFF
sample of clusters from z=0 to z=2. The solid lines are for the
deprojection methods assuming a universal temperature profile, and the
dashed are for isothermal $\beta$-model methods. It is clear that for
X-ray methods, shown in the top panel, there is very little difference
in the mass estimates from one method to another (both show a $\sim
10-12\%$ bias), but in the case of
the SZE, there is a significant difference in the distribution of mass
estimates. Not only is the scatter much smaller and more symmetric in
the UTP method, but there is almost no bias ($\sim - 4\%$)in gas mass estimates for
the full sample, whereas the isothermal method gives a 13\% bias.  This result is due to the stronger dependence on
temperature of the SZE, hence the larger effect on mass estimates of
the temperature assumption. 

\subsection{Variations with Baryonic Physics}
Lastly, we compare the mass estimates as a function of
redshift for each of the four simulation cluster
catalogs. First, in Figure \ref{bary_phys}, we see the mass estimation
for the X-ray UTP model in each of the four catalogs. The catalogs are
identified as adiabatic (AD), adiabatic plus radiative cooling (RC),
RC plus star formation (SF), SF plus thermal feedback from supernovae
(SFF). Irrespective of
the physics included, the mass estimation has similar median values
(including 5-10\% bias of the median)
and relative scatter, with the exception of the simulation which
includes radiative cooling with no other additional physics. This
results in a very high median estimate of the mass, even when
excluding the cool core of the cluster from the analysis.  
For the SZE UTP method, shown in Figure \ref{bary_phys_2}, only the
radiative cooling catalog suffers from a significant
deviation of the scatter from the other samples, but shows no strong
bias. A summary of the median values for the z=0 set of clusters in
each of the four physics samples and for four estimation methods is
shown in Table \ref{tab:bary}. It is clear from this table and Figure
\ref{bary_phys} that
the UTP methods are relatively insensitive to input physics as far as
the bias and scatter in mass estimates, except for the extreme RC case.  SZE
methods are less sensitive than X-ray methods to strong cooling, and
assuming isothermality in SZE methods is more problematic than for
X-ray methods.
\begin{table}
\caption{Median $M_{est}/M_{true}$, $r_{500}$, z=0}
\label{tab:bary}
\begin{tabular}{cccccc}
\tableline\tableline
Method & Model & AD  & RC & SF & SFF\\
\tableline
X-ray & UTP & 1.07 & 1.78 & 1.07 & 1.08 \\
SZE & UTP & 1.00 & 1.07 & 0.99 & 1.01\\
X-ray & Isothermal & 1.09 & 1.58 & 1.08 & 1.09\\
SZE & Isothermal & 1.10 & 1.30 & 1.04 & 1.08 \\
\tableline
\end{tabular}
\end{table}

The scatter does not vary significantly with redshift for any of
these methods, indicating that the gas mass estimation techniques are
relatively robust even for high redshift clusters where mergers are
more common.  However, there is slight trending of the mean and median
with redshift apparent in Figure \ref{bary_phys}. This is worrisome
for methods which assume a constant gas fraction in clusters.  Though
we can not say from this study if there is a similar trend in total
mass estimates from observables, if the trends do not match, the gas
fraction will vary spuriously as a function of redshift when measured
in this way. Using the UTP deprojection method of the X-ray surface
brightness to $r_{500}$, we find a maximum deviation in the SFF sample
from the median value of $M_{est}/M_{true}$ at z=0 of only about +3\%
at z=2. For the same sample of clusters using the SZE method, we find
a maximum deviation of -11\% at z=1. Additionally, if we use instead
the adiabatic sample of clusters, the median mass ratio from the X-ray
analysis shows a +17\% deviation from the z=0 value at z=2, though for
clusters likely detectable in X-rays (z $<$ 1), it is about +10\%. The
SZE method for adiabatic clusters shows a +9\% deviation at z=1.5.
These trends could be problematic, particularly because they appear to
depend on the assumed cluster physics. The SZE results in fact show
opposite trending, from overestimation in the adiabatic case to
underestimation in the SFF sample, when varying the physics. A full
analysis of the gas fraction measurement, including total mass
estimation from observables is necessary to understand the impact of
these trends on cosmological studies. We defer that analysis to later work.
 
The effect of gas mass overestimation in the cooling only sample of
clusters is in conflict with earlier suggestions (e.g. \citet{allen})
that cool core clusters may be good candidates for
mass estimation since it is presumed that they are more relaxed dynamically. In our samples, cool core clusters
do not qualitatively appear to be particularly relaxed systems, as has
been previously shown in \citet{motl04}.
Indeed, in our analysis of cool core clusters, the X-ray mass
estimates are biased very high on average, and have much larger
scatter than non-cool core clusters.  SZE mass estimators suffer from
a smaller bias in the median value of the sample, but have larger
scatter than estimates from non-cool core clusters. This indicates
that the effects of radiative cooling in the simulation extend beyond
the cool core region, rendering assumptions about temperature and
density profiles less applicable. Though the case of radiative cooling
only in the simulations is very extreme, and no real clusters are
likely to cool to that degree, an explanation for this apparent problem
is discussed in Section 4.4.
\subsection{Cleaning the Cluster Samples}
To more closely mimic observational studies of cluster mass
estimation, we have attempted a qualitative cleaning of the SFF cluster
sample at z=0. We have examined all the cluster projections and
removed those which have obvious double peaks in the X-ray or SZ
surface brightness images, have disturbed morphology within R =
$1h^{-1}$ Mpc, exhibit edges consistent with shocked gas, or those
that have cool cores.  The cool core clusters are identified as those
with a projected emission-weighted temperature profile which declines at small
radius, and can be removed by observers on that basis, or upon
observation of strongly peaked X-ray emission. The cool core clusters are eliminated because, as we have
already shown, they lead to strong biases in the X-ray estimates, and
increase the overall scatter of SZE estimates. While it is
straightforward to clean the simulated cluster sample, it is difficult if not
impossible to effectively clean an observed cluster sample of all non-relaxed
objects. In any case, the appearance on the sky of the cluster is not
a sufficient determinant of its dynamical state. Even with infinitely
``deep'' synthetic images, contamination of our sample still
remains. As shown in earlier sections, SZE methods are more accurate
at gas mass estimation when the sample can not be effectively cleaned
of disturbed clusters.  

\begin{figure}
\epsscale{1.2}
\plotone{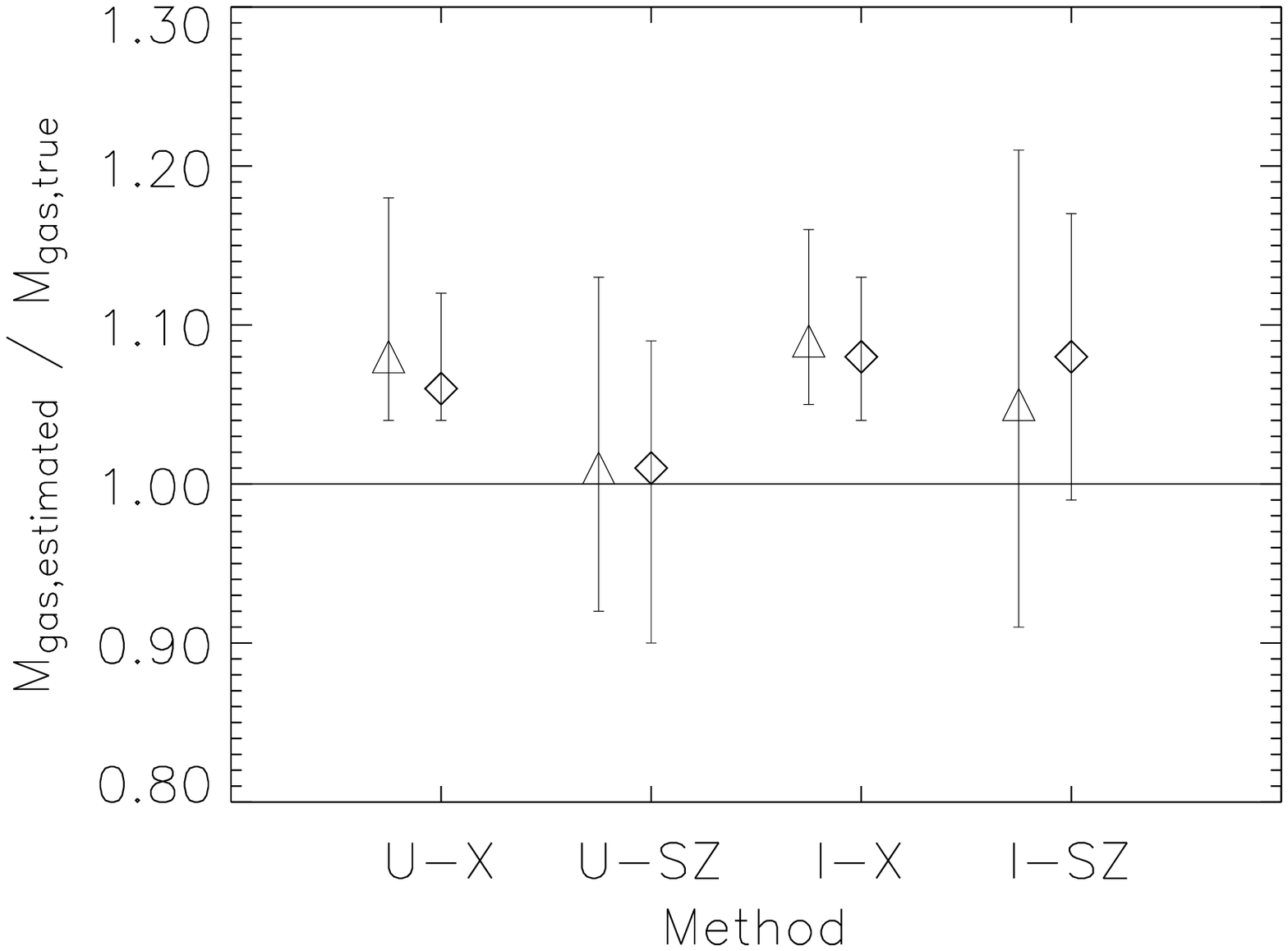}
\caption{Comparison of median values and $1\sigma$ scatter of gas mass estimates
  inside $r_{500}$ for full SFF cluster sample (triangles) and cleaned
  SFF sample (diamonds) at z=0 for each of four methods, UTP X-ray (U-X), UTP
  SZE (U-SZ), isothermal X-ray (I-X), and isothermal SZE (I-SZ) as
  described in the text}
\label{clean}
\end{figure}

Figure \ref{clean} shows for the SFF sample at z=0 the change in median value and scatter
for the mass estimates when one cleans the sample in this way.  Table
\ref{clean_tab} summarizes these results. 

\begin{table}
\caption{Scatter in $M_{est}/M_{true}$, $r_{500}$, z=0}
\label{clean_tab}
\begin{tabular}{ccccccc}
\tableline\tableline
Method & Model & Sample & Median & Mean &  +1$\sigma$ & -1$\sigma$ \\
\tableline
X-ray & UTP & Full & 1.08 & 1.11 & 1.18 & 1.04 \\
X-ray & UTP & Clean & 1.06 & 1.07 & 1.12 & 1.04 \\
SZE & UTP & Full & 1.01 & 1.02 & 1.13 & 0.92 \\
SZE & UTP & Clean & 1.01 & 1.01 & 1.09 & 0.90 \\
X-ray & Isothermal & Full & 1.09 & 2.59 & 1.16 & 1.05 \\
X-ray & Isothermal & Clean & 1.08 & 1.08 & 1.13 & 1.04 \\
SZE & Isothermal & Full & 1.05 & 1.09 & 1.21 & 0.91 \\
SZE & Isothermal & Clean & 1.08 & 1.07 & 1.17 & 0.99 \\
\tableline
\end{tabular}
\end{table}
All methods of gas mass estimation show reduced scatter upon removal of disturbed and
cooling clusters from the cluster catalogs. The reduction in
scatter for X-ray methods results almost entirely from removal of
overestimates, while the reduction is more or less symmetric in SZE methods. We believe this
difference explains the small bias of the X-ray methods. Any
clusters that appear disturbed have their gas masses preferentially
overestimated by X-ray methods. The bias remains even after this
cleaning of the sample, since we have removed only clusters which
appear obviously disturbed, some of which are also in the process of merging. The effect of boosting by
mergers has been predicted in earlier studies (e.g. \citet{roettmass,
  ricker}). 

Not only is it very difficult to determine
whether a cluster is relaxed or not from its surface brightness
profile, but even a very minor merger can result in boosting
significant enough to contaminate the mass estimation. Essentially, no clusters in the
sample match the model assumptions closely, and this fact manifests
itself as a bias in the mass estimates.  Removal of obviously disturbed clusters brings down
the mean and median values slightly, along with the high end scatter,
but does not improve the scatter at the low end. With a cleaned
sample, however, the X-ray methods
actually have smaller scatter in gas mass estimation than do SZE
methods, as illustrated in Table \ref{clean_tab}. 

\begin{figure}
\epsscale{1.2}
\plotone{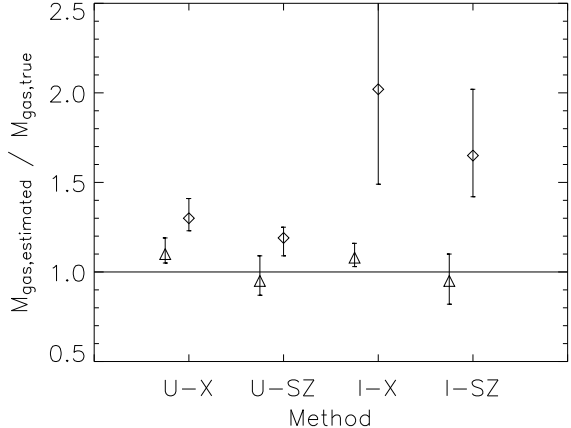}
\caption{Comparison of median values and $1\sigma$ scatter of gas mass estimates
  inside $r_{500}$ for disturbed clusters in the SFF sample
  (triangles) and cool core clusters (diamonds) at z=0 for each of
  four methods, UTP X-ray (U-X), UTP SZE (U-SZ), isothermal X-ray
  (I-X), and isothermal SZE (I-SZ) as described in the text}
\label{cool_dist}
\end{figure}

While the distribution of gas mass estimates for apparently
disturbed or merging clusters is consistent with an unbiased sample,
the disturbed clusters do show a bias in the UTP X-ray sample. In
all cases the cool core clusters show a high mass estimation bias,
which is particularly marked in the isothermal methods. This is unexpected,
since we have excised the cool core from the surface brightness
profiles in order to do the gas mass calculation. 
Figure \ref{cool_dist}
shows the median values and $1\sigma$ scatter for the mass estimates of
cooling clusters and those with disturbed morphology in each
case. We find that most of the improvement in the overall scatter comes from
the removal of the cool core clusters, and a marginal improvement
comes from removing clusters which look non-relaxed in
projection. This is particularly true in the case of the isothermal
model fits to the cluster surface brightness. In fact, for X-ray
methods, removing only cool core clusters results in nearly
identical scatter in mass estimates as a fully cleaned
sample. Therefore, it is clear that removal of clusters with apparently
disturbed morphology contributes little to reducing bias and scatter
in mass estimation in these samples. A similar result is described in
\citet{ohara} for real clusters, who find that the scatter in
observables and mass estimates from cool core clusters is higher than
for non-cool core clusters, and that clusters that appear to be
disturbed or merging do not increase scatter in those same
quantities. This result is consistent with the lack of strong trending
in the relative scatter of gas mass estimates with redshift. In fact,
it is likely given the results of this cleaning analysis that the
scatter is generally dominated by cool core clusters, since 
cool core clusters represent a roughly constant fraction with redshift
of the SFF cluster sample. 

\subsubsection{Example Clusters: Differences in SZE and X-ray methods}
In order to understand why cool core clusters should give biased mass
estimates, and the 5-10\% bias of the median values of the X-ray
estimates, we look at
some of the simulated clusters below in Figure
\ref{bias_image}. Images from three of the simulated clusters from the star
formation with feedback catalog are shown.  The top row contains images of  a mostly relaxed,
high mass cluster ($M=1.5\times10^{15}M_{\odot}$), the middle row is a cluster
that appears disturbed in morphology, with $M =
5.3\times10^{14}M_{\odot}$, and the bottom row is a lower mass ($M=1.4\times10^{14}M_{\odot}$),
apparently relaxed cluster with a very cool core. The columns show from
left to right, projected X-ray luminosity, projected Compton
y-parameter, and projected emission weighted temperature. The middle
row cluster has a clear double-peaked structure in both X-rays and SZE
images, and would be classified as a merging cluster purely based on
appearance. 

Table \ref{tab:three_clust} shows the gas mass estimation for
geometric deprojections inside $r_{500}$ assuming a UTP model for temperature for both
X-ray and SZE synthetic observations of the three clusters shown in
Figure \ref{bias_image}. For the relaxed cluster, both methods do a
reasonable job at measuring the gas mass in the cluster accurately,
though the SZE method is about a 6\% underestimate.  In the disturbed
morphology case, the X-ray method generates a more than 50\%
overestimate of the cluster gas mass, while the SZE shows just a 7\%
overestimate. In the cluster where the gas is strongly cooled in the
core, even though we have excluded the cold core from the analysis,
again the X-ray estimate is much higher than that computed from 
SZE information.  These examples indicate a common trend throughout
the data, when cluster masses tend to be overestimated by
observations (\eg  when the emission is boosted by mergers or cooling)
the X-ray is more strongly affected than the SZE emission.  This is
expected and results from the X-ray emission having an $n_e^2$
dependence, thus being dominated by boosting effects in the cluster core. 

\begin{figure}
\epsscale{1.2}
\plotone{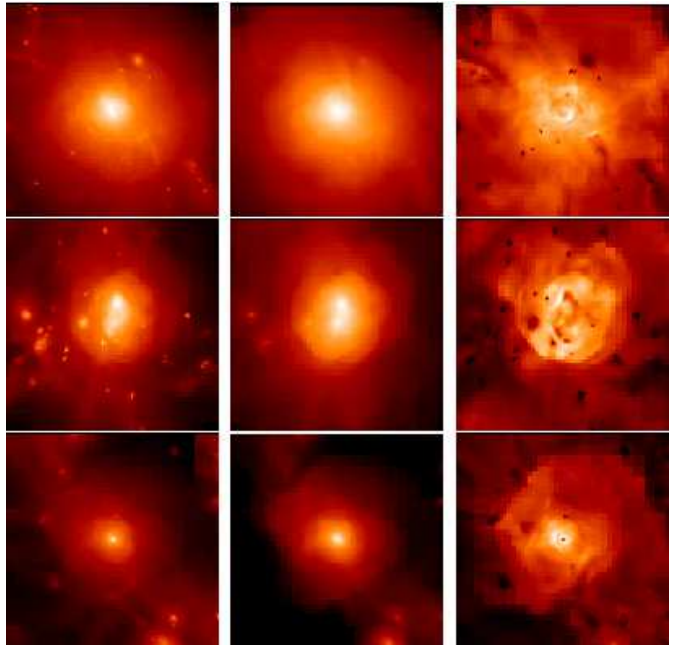}
\caption{Three clusters.  Top row appears relaxed, middle row has
  disturbed morphology and bottom row has a cool core. Left column
  shows X-ray surface brightness from simulated cluster, middle column
  is Compton $y$ parameter and right column is emission weighted
  temperature. Images are $\sim 5 h^{-1} Mpc$ on a side.} 
  \label{bias_image}
\end{figure}

\begin{table}
\caption{Gas Mass Estimates for Three Clusters}
\label{tab:three_clust}
\begin{tabular}{cccc}
\tableline
\tableline
State & X-ray $M_{est}/M_{true}$ & SZE $M_{est}/M_{true}$ & $M_{tot}$ ($M_{\odot}$)\\
\tableline
Relaxed & 1.00 & 0.94 & $1.5\times10^{15}$ \\
Disturbed & 1.54 & 1.07 & $5.3\times10^{14}$ \\
Cooling & 1.39 & 1.10 & $1.4\times10^{14}$ \\
\tableline
\end{tabular}
\end{table} 

\subsubsection{The Problem with Weighing Cool Core Clusters}
\begin{figure}
\epsscale{1.2}\plotone{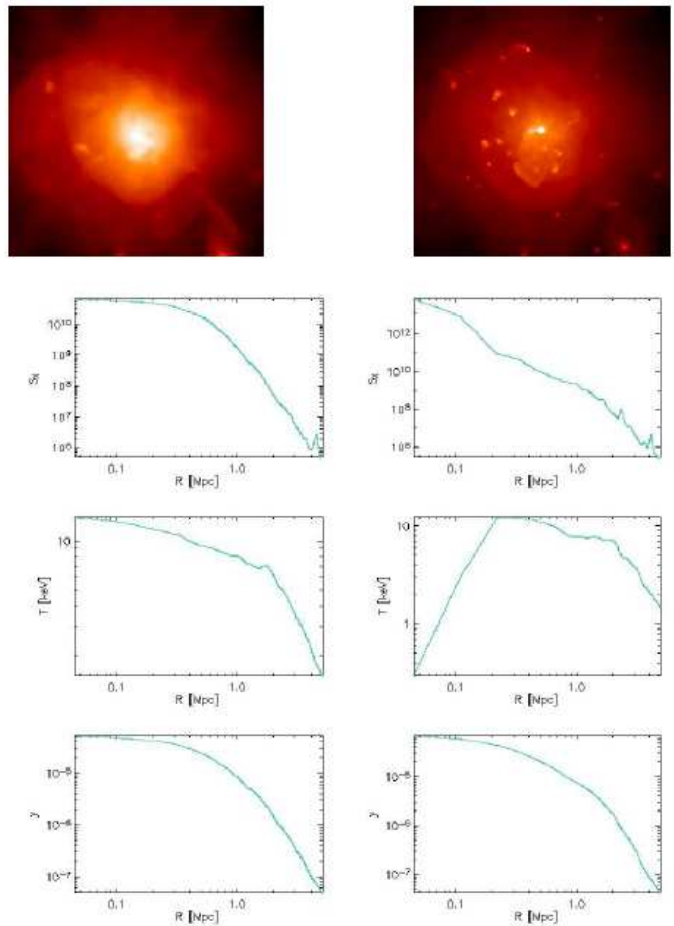}
\caption{Same cluster from different cluster catalogs. Left column is from
  adiabatic physics only simulation, right column is from catalog run
  with addition of radiative cooling. Top row are images of X-ray surface brightness, second row are X-ray surface brightness radial profiles,
  third row are radial profiles of projected emission-weighted
  temperature, bottom row are radial Compton $y$ parameter profiles.  Images are $\sim 5 h^{-1} Mpc$ on a side.} 
  \label{ad_cool}
\end{figure}

Though it is clear the SZE is less affected than the X-ray when
calculating gas mass for the cool core clusters in the simulation, it
is still not obvious why there should be a bias at all.  A look at the
cluster projected radial profiles helps to illustrate the difference
between cool core and non-cool core clusters.  Figure \ref{ad_cool} shows
images of X-ray surface brightness, and radial
profiles of X-ray surface brightness, SZE, and projected
emission-weighted temperature of the same cluster from two different
simulations. One was run with only adiabatic physics and the other
with radiative cooling turned on, otherwise the simulations are
identical. Where the Compton y-parameter profiles are nearly identical
outside the cool core ($\sim 200 \; kpc$), the X-ray surface brightness
profiles are significantly different.  In the adiabatic sample, the surface
brightness has a well defined core, and the slope breaks outside of
the core. In the cool core case, the radial X-ray surface brightness
profile has a very different appearance.  Fitting a $\beta$-model to
this profile overestimates the density profile, and hence the mass. It is apparent that the result of
radiative cooling in the simulation is to impact the cluster in
regions outside the cool core as well as inside. Whether heating
effects in real clusters will offset such deviations remains to be seen.

\begin{figure}
\epsscale{1.2}
\plotone{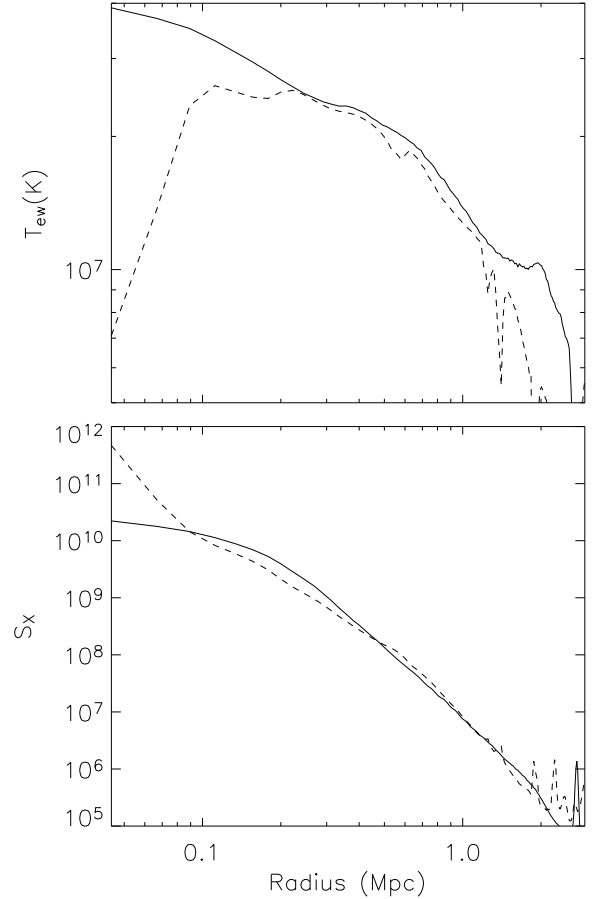}
\caption{Top: Emission weighted projected temperature radial profiles
  for two clusters of nearly equal mass at z=0 in the SFF
  sample. Solid line is for a cluster without a cool core, dashed is a
  cluster with a cool core. Bottom: Surface brightness profiles of
  same two clusters.}
  \label{comp2}
  \epsscale{1.0}
\end{figure}

For a more direct comparison, Figure \ref{comp2} shows
the radial temperature and surface brightness profiles of two nearly
equal mass clusters from the SFF simulation volume.  The main difference is
that one has a cool core at z=0, and one does not. It is obvious that
the cool core extends to roughly 200 kpc in this cluster as well as
the one in Figure \ref{ad_cool}. The profiles of surface brightness
are quite different, the non-cool core (NCC) cluster (solid line) has a
profile shape well approximated by a $\beta$-model, while the cool
core (CC) cluster clearly does not.  Table \ref{tab:mratio} shows the results of
fitting a $\beta$-model to each of two projections of each of these
clusters. For the NCC cluster, we get close to the
correct gas mass, the numbers are slightly high, consistent with the
full sample of X-ray clusters.  The CC cluster has
significantly biased gas mass estimates, a result of the 
$\beta$-model fit to the profile. Several indications exist
identifying the $\beta$-model fitting as the problem, one of which was
illustrated earlier in Figure \ref{cool_dist}. The bias in gas mass
estimates for UTP deprojection models is slightly higher than for the
full sample of clusters, the bias is much larger for the isothermal
$\beta$-model fits. In fact for the clusters in Table
\ref{tab:mratio}, the UTP X-ray method overestimates a gas mass only
by about 20\%.  

\begin{figure}
\epsscale{1.2}
\plotone{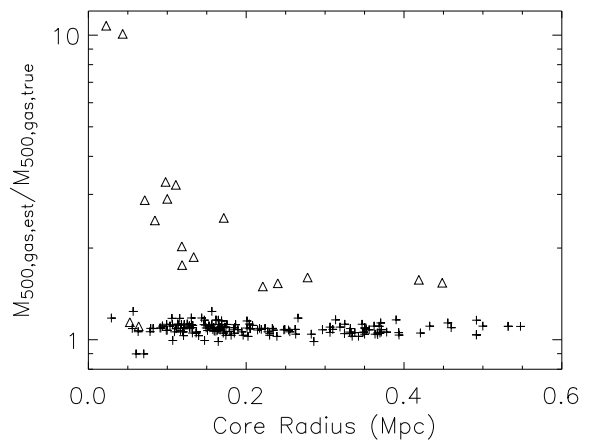}
\caption{Ratio of estimated to true gas mass of clusters plotted against fitted value
  of the $\beta$-model core radius for the SFF sample of clusters at
  z=0 with a fitting radius of $r_{500}$. Cool core clusters are
  represented by triangles. Cool core clusters show consistent
  overestimates in gas mass from X-ray fitting and the magnitude of
  the overestimate is correlated to the size of the core radius.}
  \label{mratio}
\end{figure}

A first indication of differences is that CC clusters fit to
systematically lower values of the core radius than do NCC
clusters, and for the CC clusters, the small core radius fit
correlates to higher mass overestimates. Table \ref{compcool} shows
the comparison of median fitted values for $\beta$-model parameters
for all clusters at z=0, separated into CC and NCC
samples. Figure \ref{mratio} shows the correlation between the ratio
of estimated to true mass for the clusters plotted against the value
of the fitted core radius.  There is a wide distribution of core
radii for NCC clusters, but fitting CC clusters typically results in small
core radius values. There appears to also be some correlation between
smaller core radii and larger mass overestimates for CC clusters. These effects are indications that fitting the
$\beta$-model to a CC cluster's surface brightness profile is
problematic. It is clear that the shape of the surface brightness
profile is significantly different in CC clusters than in
NCC clusters. 

\begin{figure*}
\plotone{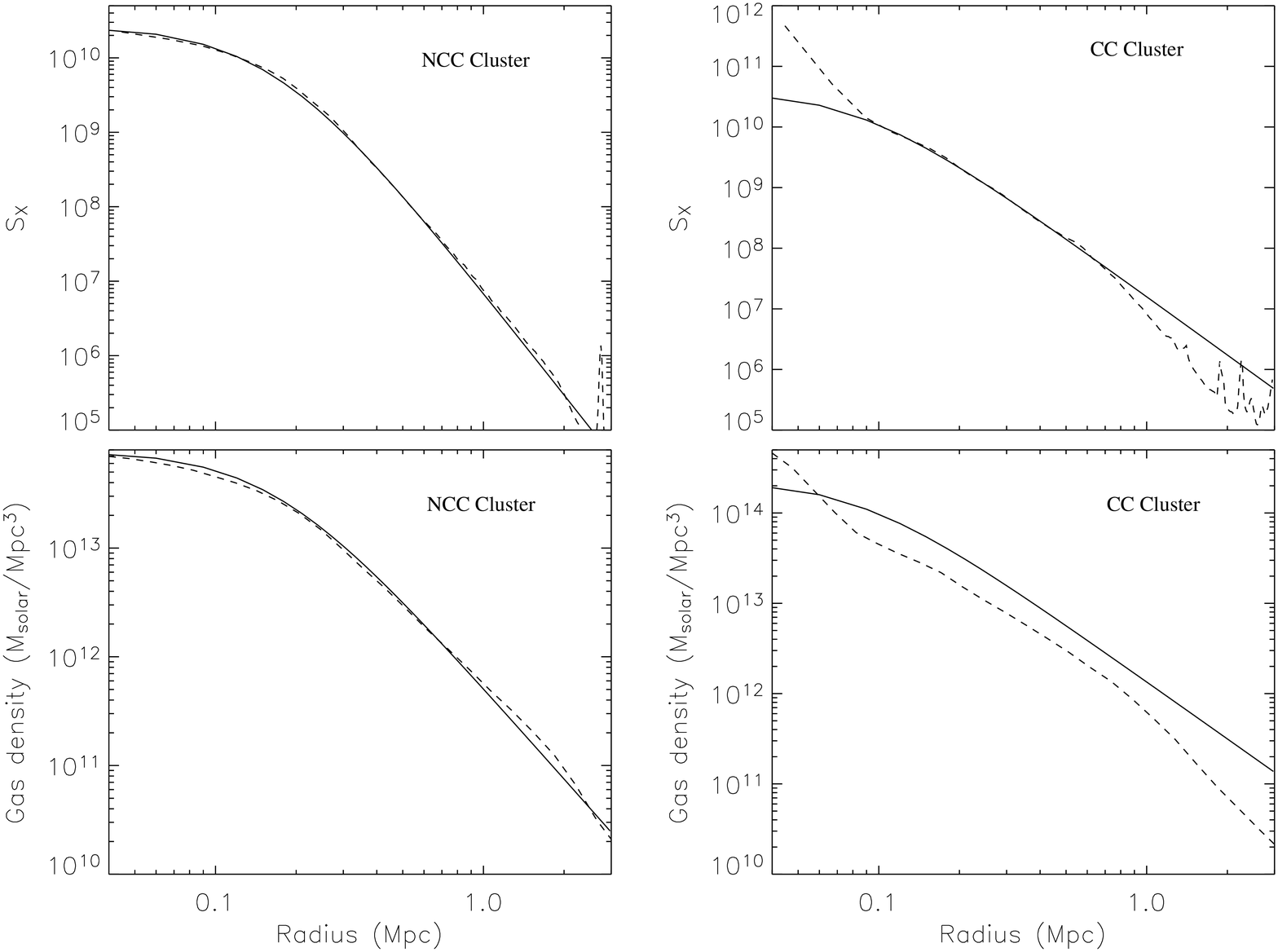}
\caption{Top Left: X-ray surface brightness profile for non-cool core
  cluster (NCC) described in text overlaid with best fit
  $\beta$-model. Top right: X-ray surface brightness profile and best
  fit model for cool core cluster. Bottom left: True gas density
  profile of the non-cool core cluster in above panel, again overlaid
  with best fit $\beta$-model from the surface brightness. Bottom
  right: same as bottom left, but for cool core cluster whose surface
  brightness is shown in above panel.}
  \label{sxvstew}
\end{figure*}

For the two clusters whose profiles are shown in Figure \ref{comp2},
we show the comparison of the $\beta$-model fits in Figure
\ref{sxvstew}. The top panels are for the fit to the X-ray surface
brightness in each case, where the CC cluster fit applies to the
region outside the cool core. While both models (solid line) appear to
fit the data fairly well in the radius regime of interest, the lower
panels address the true problem.  The $\beta$-models which correspond
to the upper panels are shown compared to the true density profile of
each cluster. For the NCC case, the model is a good fit, leading to a
good mass estimate for the cluster.  In the CC case, the model
strongly overestimates the density profile, leading the the
overestimate in mass. This effect is very typical in cool core
clusters in our sample.  Clearly there is a breakdown in the
assumptions of the $\beta$-model for cool core clusters.  The physical
reasons for this discrepancy are not obvious, and are likely
complicated. One expects, however, that if there is more gas
contributing to the surface brightness with temperatures below the
assumed isothermal temperature in CC clusters that may lead to
overestimates in the density. This explanation may help explain the
bias in UTP method gas mass estimates. It is clear that directly
deprojecting the X-ray and SZE emission from CC clusters also results
in somewhat biased estimates. We plan further investigations of cool core
clusters in future studies. 
\begin{table}
\caption{Comparison of Equal Mass Clusters, SFF}
\label{tab:mratio}
\begin{tabular}{cccccc}
\tableline
\tableline
Cluster & Proj & $M_{est}/M_{true}$ & $r_{c}$ (Mpc) &
$\beta$ & True $M_{tot}$ ($M_{\odot}$)\\
\tableline
CC & 1 & 2.46 & 0.084 & 0.70 & $1.91\times 10^{14}$ \\
CC & 2 & 2.87 & 0.071 & 0.69 & $1.91\times 10^{14}$ \\
NCC & 1 & 1.16 & 0.17 & 0.93 & $1.94\times 10^{14}$ \\
NCC & 2 & 1.08 & 0.15 & 0.92 & $1.94\times 10^{14}$ \\
\tableline
\end{tabular}
\end{table} 
\begin{table}
\caption{Comparison of X-ray $\beta$-model Parameters, z=0, $r_{500}$}
\label{compcool}
\begin{tabular}{cccc}
\tableline
\tableline
Cluster Sample & $\langle S_{X0}\rangle$ & $\langle r_c \rangle$ &
$\langle \beta \rangle$\\
\tableline
CC & $4.42\times10^{10}$ & 0.12 & 0.77 \\
Non Cool Core & $1.41\times10^{10}$ & 0.18 & 0.80 \\
\tableline
\end{tabular}
\end{table} 

In cases where the gas is strongly cooled, or has disturbed
morphology, the X-ray gas mass determinations are more strongly
overestimated.  While SZE methods show bias for individual clusters,
it is generally smaller than for the X-ray
methods.  The stronger density
dependence of the X-ray emission, and its resultant boosting from
density enhancements from cooling, mergers and substructure apparently is
responsible for the bias persisting even using profiles to large
radii. Any effects which enhance density in localized areas of the
cluster will tend to cause overestimates in the cluster total gas mass when
using X-ray methods.  We expect that X-ray
methods of mass estimation for cluster gas will result in a slight
bias (5-10\%). 

\subsection{Prospects for Cluster High Resolution SZE Observations}
A growing number of SZE telescopes either are active now, or are
expected to come online in the next few years. The largest of the
single dish instruments capable of resolving radial SZE
profiles for many clusters, will be the Large Millimeter Telescope (LMT).
Cluster radial SZE profiles will also be generated with the Atacama
Large Millimeter Array (ALMA) in a few years time. With ALMA and the LMT, we expect to
be able to measure cluster SZE profiles to relatively large radii. In
that case, gas mass estimation using SZE profiles can be done to
fairly high accuracy.  Cluster X-ray profiles are available for many
clusters out to $r_{500}$ with \textit{XMM-Newton} and
\textit{Chandra} imaging. We expect the LMT to observe cluster profiles to
similar or larger radii. 

In order to estimate the expected cluster radial
coverage for the LMT using the Astronomical Thermal Emission Camera (AzTEC) bolometer array, we use the minimum detectable flux density calculation from
\citet{pahol}
\begin{equation}
F_{min} = \frac{\sqrt{2}k_B T_{sys}}{A_d \sqrt{t \Delta \nu }},
\end{equation}
where $T_{sys} = 54K$, and
$\Delta \nu = 42 \; GHz$ is an estimate for the AzTEC band centered at
144 GHz, and $A_d = \pi (50/2)^2 m^2$.
This gives a minimum detectable flux density for the LMT of
\begin{equation}
F_{min} = 0.26 \frac{1}{\sqrt{t}} \; mJy.
\end{equation}
To calculate the flux density at $r_{500}$ and $r_{200}$ for simulated clusters,
we determine that the typical radial profile average value for
clusters of $y(r_{500}) \approx 10^{-8}$ and $y(r_{200}) \approx
10^{-9}$. Calculating the change in intensity of the CMB at 144 GHz,
we estimate
\begin{equation}
\Delta S_{r_{500}} = 2.53 \times 10^{-7} \;mJy/arcsec^{2}.
\end{equation}
The value at $r_{200}$ then is a factor of 10 lower. At $r_{500} (\approx 1h^{-1} Mpc)$ set at z=0.1 in a ring $10
\arcsec$ wide, the total flux density from that ring will be F = $12.5 \mu
Jy$. For $r_{200} (\approx 1.5h^{-1} Mpc)$ at z=0.1, the flux density is F =
$1.87 \mu Jy$. To observe these with the LMT would require $t \approx
433s$ and $t \approx 72$ minutes per pointing, respectively. These times are very short compared to a typical, high 
quality,  X-ray observation of $t \approx 50 ks \approx 14 hours$. While this estimate is
probably generous, considering the difficulties of atmospheric
subtraction and the removal of other instrumental effects, observation
of cluster SZE signal out to $r_{200}$ appears to be possible with
the LMT.       
\begin{table}
\caption{Rating of Mass Estimates with Cleaned Samples}
\label{final_tab}
\begin{tabular}{ccccccc}
\tableline
\tableline
Method & Model & Radius & Median & Mean &  +1$\sigma$ & -1$\sigma$ \\
\tableline
Xray & UTP & $r_{500}$ & 1.06 & 1.07 & 1.12 & 1.04\\
Xray & Iso & $r_{500}$ & 1.08 & 1.08 & 1.13 & 1.04 \\
SZE & UTP & $r_{200}$ & 0.98 & 0.99 & 1.04 & 0.94 \\
Joint & Geometric & $r_{500}$ & 1.09 & 1.09 & 1.15 & 1.04 \\
SZE & y-M & $r_{500}$ & 0.96 & 0.98 & 1.09 & 0.92 \\
SZE & UTP & $r_{500}$ & 1.01 & 1.01 & 1.09 & 0.90 \\
SZE & Iso & $r_{500}$ & 1.08 & 1.07 & 1.17 & 0.99 \\
Xray & $T_X$-M & $r_{500}$ & 1.03 & 1.01 & 1.24 & 0.79 \\
\tableline
\end{tabular}
\end{table}  
SZE with UTP methods out to large cluster radius produce high
precision mass estimates, even without filtering the sample for
merging or disturbed clusters. Therefore, from a standpoint of accuracy and simplicity,
SZE methods appear to be excellent tools to generate reliable cluster mass
measurements to enable precision cosmology with clusters of galaxies.


\section{Conclusions} 
The highest precision to which cluster gas masses can be measured using
	 typical assumptions and either X-ray, SZE or a combination of
	 both measurements is $\sim10$\% to $1\sigma$ confidence. This
	 study does not include instrumental or other observational
	 effects, and so is an indication of the limiting ability of
	 observations to correctly gauge the mass of clusters in the
	 ideal case. Our study suggests that when using X-ray observations
	 in a fixed band the scatter is likely higher by a factor of
	 2-3. These limits in precision are a direct result of
	 the deviation of the simulated clusters from simple
	 assumptions about their physical and thermodynamic properties, dynamical state,
	 and sphericity. To summarize our results, we include Table \ref{final_tab}, which shows the
relative error for the best cluster gas mass methods compared to one
another in order of limiting accuracy.  Table \ref{final_tab} also includes scatter estimates for
SZE out to $r_{200}$, as well as scatter in total mass estimates using
cleaned samples for $y_{500}$ versus mass and the cluster $T_X-M$
relation as calculated in \citet{motl05}. 
	 
	SZE methods of gas mass estimation assuming a universal
temperature profile in the cluster gas produce the smallest scatter
when estimating masses in a raw sample of clusters.  Cleaning the
cluster sample for disturbed or merging clusters is much less
important in SZE methods, particularly when profiles can be observed
to large radius. As a practical matter, SZE methods are superior for
mass estimation for large samples of clusters out to high
redshift. This is consistent with previous work, which shows that
cluster SZE observations to large cluster radius smooth out boosting
effects in the cluster core, and therefore is more representative of
the true cluster potential than the X-ray emission. 

	 X-ray mass estimation methods using radial profile fitting do
	 slightly better than SZE methods as far as relative scatter
when using a sample cleaned of obvious mergers and disturbed morphology
clusters.
	 However, X-ray methods show a 5-10\% bias in median values
	 which is absent is SZE methods. The bias is a result
	 of substructure and mergers in clusters, events which enhance
	 the gas density locally. The stronger density dependence of
	 X-ray emissivity tends to drive up the X-ray mass estimates
	 more than the SZE estimates. 

	 Mass estimates fitting radial profiles to a universal
	 temperature profile (UTP) have smaller scatter than similar
	 estimates assuming isothermality, particularly for SZE
	 methods.  Also, the accuracy of X-ray mass estimates improves only marginally by
increasing the available radius of the observational profiles. This
	 effect can be explained by the strong dependence of the X-ray emission on density, thus on the
properties of the cluster core.   SZE methods however improve dramatically with profile data
out to higher radii, provided one uses a UTP and does not assume the
cluster to be isothermal. This is expected since UTP models are a better fit
	 to the real temperature distribution in clusters than
	 isothermal models, and the SZE depends more strongly on
	 temperature than does the X-ray emission.   

While the scatter in these estimates varies from method to method, in
principle this could be overcome by using a large cluster
sample. However, the bias can not be removed in this
way. Conveniently, the very small variation in gas mass estimate bias when varying
baryonic physics in the simulation suggests that one can use the
simulations to correct the bias in observationally determined gas
masses. In all but the most extreme case (radiative cooling only) of
our samples, the bias in X-ray estimated gas masses is about
7-9\%. The bias is also the same (9\%) when using a soft (0.5-2.0keV) band
and spectroscopic-like temperature to estimate masses.  This indicates
that the bias is relatively robust, and can be corrected. 

Cool core clusters in our catalogs are particularly poor candidates for precision mass
estimation in disagreement with previous assumptions.  Even when excising
the cool core from the analysis, mass estimation shows larger
scatter in both X-ray and SZE methods than for non-cool core
clusters. X-ray methods also generate a very high
($\sim50-100\%$) bias in the median value of the cluster gas mass.
The proximate cause of bias in cool core clusters is the poor fit of
the $\beta$-model from the X-ray surface brightness profiles to the
cluster's true gas density profile. The shape
of the surface brightness profile in cool core clusters is markedly
different from the profile in non-cool core clusters, and is similar
throughout the cool core samples of clusters. This results in
$\beta$-model parameters which have systematically small values of the
core radius, and large central surface brightness values. When using
UTP deprojection methods, the bias and scatter in cool core cluster
gas mass estimates is greatly reduced. This remaining bias may also be
due to a larger amount of cooler gas contributing to the X-ray surface
brightness. This difference in cool gas content is possibly related to
differences in merging history between CC and NCC clusters.

In order for this study to provide direct guidance to observers, it
remains to include instrumental effects. To truly
gauge the likely accuracy of gas mass estimates, we need to examine
these effects in detail by simulating the instrumentation and
background. Additionally, it is important to examine the systematic
errors generated when computing cluster total masses from either the
assumption of hydrostatic equilibrium in the cluster, or from lensing
data. This is left to future work.    

\acknowledgments The simulations presented in this paper were
conducted at the National Center for Supercomputing Applications at
the University of Illinois, Urbana-Champaign. We wish to acknowledge
the support of the National Science Foundation through grant
AST-0407368. We acknowledge the support of the \textit{Chandra} X-ray
Center and NASA through grant TM3-4008A. We thank an anonymous referee
for very useful and extensive comments. We also wish to thank the
Laboratory for Computational Astrophysics for support of the
\textit{Enzo} code.
%===============================================================================
% References
%===============================================================================
\bibliography{ms.bib}
%===============================================================================

% The End
%===============================================================================

\end{document}